\documentclass{jfm}

\usepackage{natbib}
\usepackage{graphicx}
\usepackage{epstopdf, epsfig}

\usepackage[utf8]{inputenc}
\usepackage[T1]{fontenc}

\usepackage{placeins}
\usepackage{float}
\usepackage{xcolor}

\usepackage{multirow}
\usepackage{mathrsfs}
\usepackage{amsmath} 
\usepackage{amssymb}

\usepackage{bm} 
\usepackage[colorlinks=false,hidelinks]{hyperref}  

\title{Turbulence induced by a swarm of rising bubbles from coarse-grained simulations}
\shorttitle{Turbulence induced by a swarm of rising bubbles}

\author{
R\'emi Zamansky\aff{1}\corresp{\email{remi.zamansky@imft.fr}}, 
Florian Le Roy De Bonneville\aff{1}\footnote{Present address: Institute of Meteorology and Climate Research, Institute for Hydromechanics, Karlsruhe Institute of Technology, Germany}, 
and Fr\'ed\'eric Risso\aff{1} }
\shortauthor{R. Zamansky, F. Le Roy De Bonneville,  and F. Risso}

\affiliation{\aff{1}
Institut de Mécanique des Fluides de Toulouse (IMFT), Université de Toulouse, INPT, UPS, CNRS, Toulouse, France
}

\begin{document}

\maketitle

\begin{abstract}

We performed numerical simulations of a homogeneous swarm of bubbles rising at large Reynolds number, $Re=760$ with volume fractions ranging from 1\% to 10\%. We consider a simplified model in which the interfaces are not resolved, but which allows us to simulate flows with a large number of bubbles and to emphasize the interactions between bubble wakes. The liquid phase is described by solving, on an Eulerian grid, the Navier-Stokes equations, including sources of momentum which model the effect of the bubbles. The dynamics of each bubble is determined within the Lagrangian framework by solving an equation of motion involving the hydrodynamic forces exerted by the fluid accounting for the correction of the fictitious self-interaction of a bubble with its own wake. The comparison with experiments shows that this coarse-grained simulations approach can reliably describe the dynamics of the resolved flow scales. We use conditional averaging to characterize the mean bubble wakes and obtain in particular the typical shear imposed by the rising bubbles. On the basis of the spectral decomposition of the energy budget, we observe that the flow is dominated by production at large scales and by dissipation at small scales and we rule out the presence of an intermediate range in which the production and dissipation are locally in balance. We propose that the $k^{-3}$ subrange of the energy spectra results from the mean shear rate imposed by the bubbles, which controls the rate of return to isotropy.

\end{abstract}


\section{Introduction}

In this paper, we are interested in the flow induced by the rise of a swarm of bubbles.
In the configuration considered here, the only source of momentum is the buoyancy acting on the bubbles, and without the bubbles the liquid would remain at rest. 
It is a complex system in which the movements of the bubbles and the liquid are coupled, leading to the emergence of collective phenomena and original properties of the flow \citep{Risso:2018}. 
We consider the case of a homogeneous swarm of large deformed bubbles, with a Reynolds number, based on the bubble size and terminal velocity $v_0$, of a few hundreds, so that each bubble generates an intense wake.

A first manifestation of collective effects is the decrease in the average bubble rising speed as the gas volume fraction $\alpha$ increases \citep{Zenit:2001,Garnier:2002,Riboux:2010}.
On the other hand, the main cause of bubble velocity fluctuations is attributed to wake instabilities. 
Indeed, when the deformation of a bubble and its Reynolds number are large enough, the wake becomes unstable and the bubbles exhibit path oscillations \citep{Mougin:2001,Zenit:2008,Ern:2012} and this seems to remain the case even for high volume fractions, of the order of 30\% \citep{Colombet:2015}. 
These wake-induced fluctuations are probably the reason why bubbly flows can remain homogeneous, and be generated in laboratory bubble columns \citep{Wijngaarden:2005}.
However, the stability of homogeneous bubble columns remains an open problem and is limited to reasonably small geometries (of the order of one meter) with well-controlled uniform bubble injection.
In most industrial applications, the gas volume fraction is not homogeneous throughout the flow and large-scale buoyancy-induced motions develop \citep{Mudde:2005}.

Fluid fluctuations exhibit very specific properties that have been identified experimentally \citep{Lance:1991,Zenit:2001,Garnier:2002,Risso:2002,Rensen:2005,Martinez-Mercado:2007,Riboux:2010,Mendez-Diaz:2013,Prakash:2016,Almeras:2017}. 
Several contributions to the fluid fluctuations can be distinguished  \citep{Risso:2018}.
For a homogeneous swarm of bubbles, there are, on the one hand, the localized perturbations around the bubbles (due to both potential effects and their direct wake) and the turbulence induced by the bubbles.
The latter is essentially driven by the interactions between the bubble wakes \citep{Riboux:2010,Riboux:2013,Amoura:2017,Risso:2018}. 
The mean kinetic energy varies approximately as $K \sim \alpha v_0^2$.
The velocity fluctuations are strongly anisotropic, with the variance of the vertical velocity being more intense than that of the horizontal velocity. 
Their Probability Density Functions (PDFs) are non-Gaussian, with exponential tails and a strong asymmetry between the upward and downward directions.

The structure of this flow is also characteristic, and the velocity spectrum exhibits a rapid $k^{-3}$ decay in a wavenumber range extending around the bubble diameter \citep{Lance:1991,Riboux:2010,Almeras:2017}.
The origin of such a scaling law as well as its precise limits in the spectral domain remain poorly understood.
From a dimensional point of view, we can write that the energy spectrum must be written as
\begin{equation}
	E(k) = f^2 k^{-3}
\end{equation} 
 where $f$ is the inverse of a timescale. 
 \citet{Lance:1991} have proposed that the $k^{-3}$ regime is associated with an equilibrium between production and dissipation and that this frequency results from the characteristic shear rate of the wakes.
 Other flows also present a $k^{-3}$ spectrum. 
 This is the case, for example, of two-dimensional turbulence at scales smaller than the energy injection scale. 
 In this flow, the flow timescale is imposed by the conservation of the enstrophy \citep{Kraichnan:1967,Batchelor:1969}. 
 Decaying turbulence subjected to intense rotation also develops a $k^{-3}$ spectrum with the timescale imposed by the rotation rate \citep{Bellet:2006}. 
 Another example concerns the turbulence under the wave surface and this time the timescale results from the frequency imposed by the swell \citep{Magnaudet:1995b,Thais:1995}.

In \S \ref{sec:model}, we present the numerical approach and the physical parameters used to simulate the flow that is subsequently analyzed.
Detailed comparisons between the numerical simulations and experimental results are presented in \S \ref{sec:exp}.
In \S \ref{sec:scales}, we proposed characteristic scales of the flow based, in particular, on the properties of the mean wakes.  
Finally, to study the mechanisms underlying the $k^{-3}$ regime, we consider the spectral decomposition of the energy budget in \S \ref{sec:spec}, and we characterize the scale-by-scale anisotropy of the flow in \S \ref{sec:aniso}.

  \section{Simulation of the bubble swarm}\label{sec:model}

Although the equations describing precisely this type of flow are relatively well known, their numerical simulation remains out of reach, due to the large spectrum of temporal and spatial scales involved.
The smallest scales are a priori associated with the interfacial dynamics and the development of a very thin boundary layer around the bubbles, while the largest scales are related to the length of the wakes and the evolution of the collective dynamics of the flow that takes place.
In order to simulate these flows, we use the approach proposed by \citet{Le-Roy-De-Bonneville:2021}.
This approach  abandons the precise description of the flow around the bubbles as well as the capillary effects while keeping a realistic dynamic of the downstream part of each wake and enables a straightforward analysis of the  structure of the liquid velocity field and the dynamics of the bubble swarm.
This modeling, based on the Euler-Lagrange approach, allows us to simulate flows with a large number of bubbles and to focus on the interactions between wakes. 
As we briefly recall below, the main difficulty of this type of calculation comes from the self-interaction of a bubble with its own wake. 
\citet{Le-Roy-De-Bonneville:2021} proposed a method enabling taking into account this effect and to accurately calculate  the trajectory of each bubble.  
This method allowed us to obtain numerical simulations of the turbulence induced by a swarm of bubbles  as illustrated in figure 1. We'll show later, in section 3, that the flow structure predicted by this approach is in good agreement with experiments.

  \begin{figure}
    \centering
	\includegraphics[width=0.49\textwidth,height=0.35\textheight, keepaspectratio]{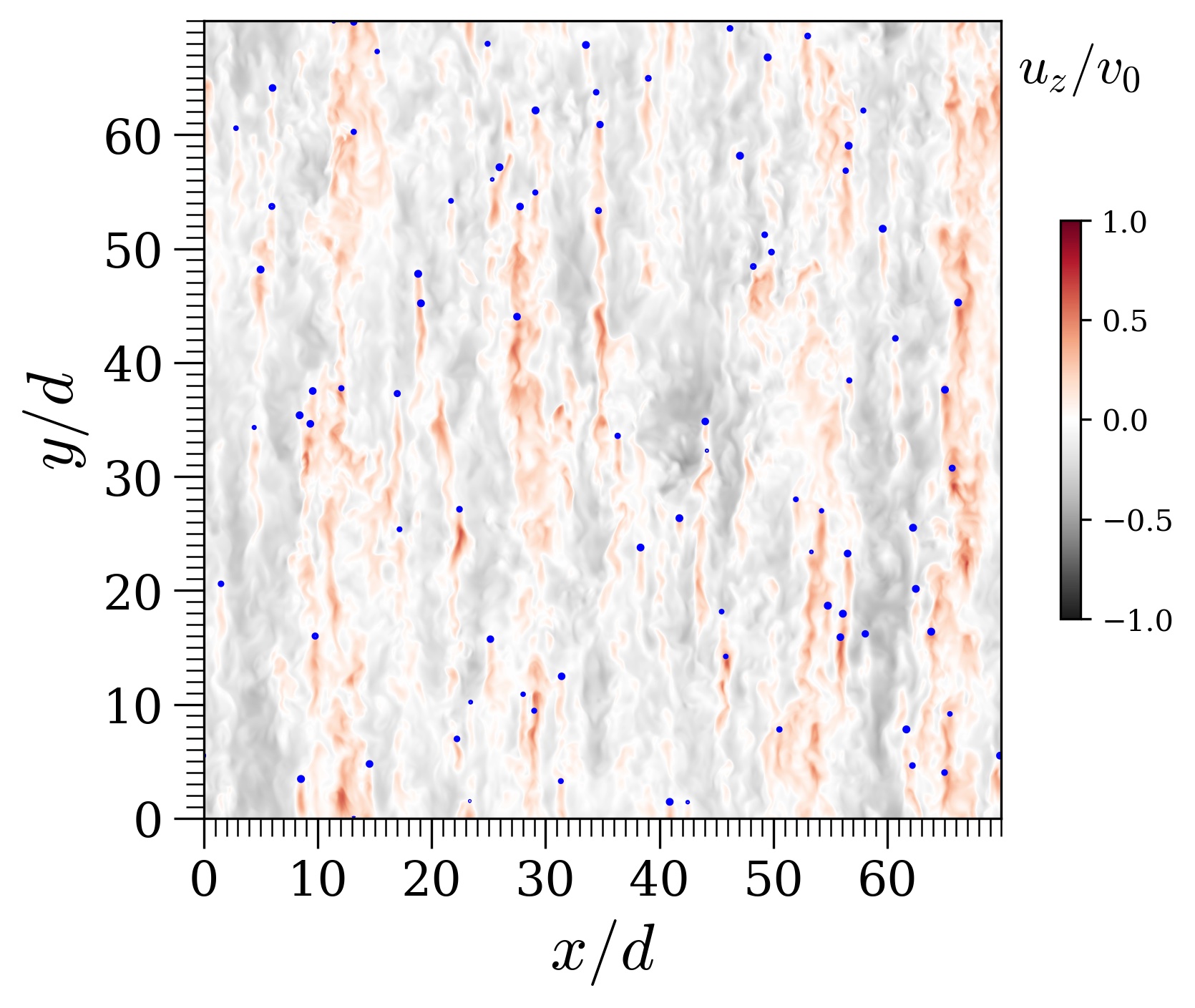}
	 \includegraphics[width=0.49\textwidth,height=0.35\textheight, keepaspectratio]{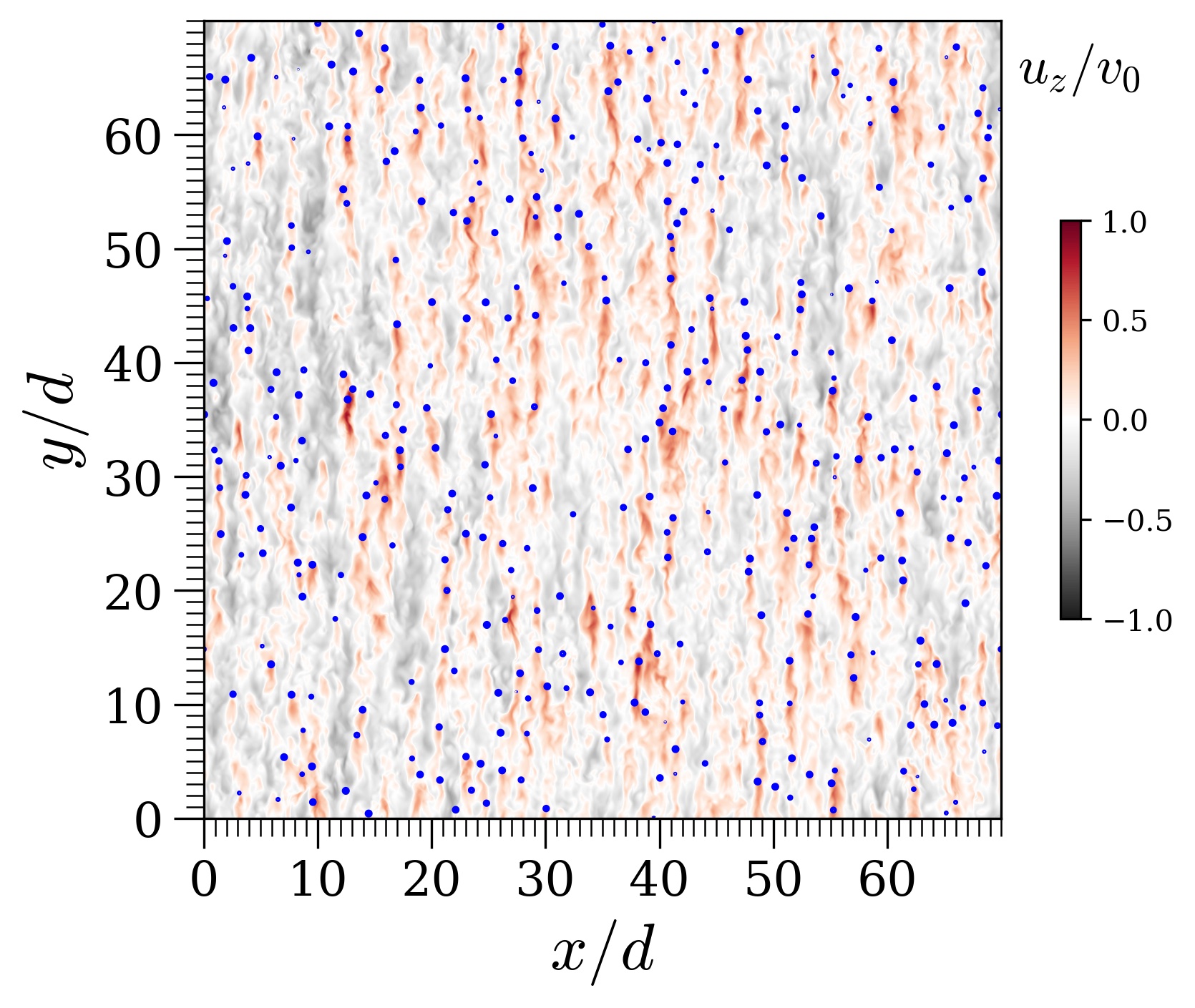}
    \caption
    {
	Snapshots of the vertical component of the liquid velocity field in a vertical plane for $\alpha=2\%$ and $\alpha=10\%$. The blue points represent the position of the bubbles.
  	 }
  \label{fig:bubble_swarm_visu}
  \end{figure}
  
\subsection{Modeling}

The use of the Euler-Lagrange approach amounts to considering a filtering of the flow field near the bubbles.
In this approach, the action of the dispersed phase on the flow is introduced as a volume source of momentum localized around the bubbles. 
The liquid velocity is given by the Navier-Stokes equations:
\begin{equation}
	D_t \bm{u}_f = -\dfrac{1}{\rho_f}\nabla p_f+\nu \Delta \bm{u}_f + \dfrac{\bm{f}}{\rho_f}  \ \ ; \ \ \nabla . \bm{u}_f = 0 \ .
	\label{eq:NS_big_bubbles}
\end{equation}
where $\bm{u}_f$ represents a filtered velocity field around the bubbles, $\nu$ the kinematic viscosity of the liquid and $\rho_f$ its density. Note that $p_f$ is the pressure variation relative to the hydrostatic pressure $P_h$ with $-1/\rho_f \nabla P_h = \langle \bm{f} \rangle$, which ensures that the computation is carried out in the frame where the average fluid velocity is zero.
These equations are supplemented by tri-periodic boundary conditions and their numerical solution are obtained by a spectral method as detailed in \citet{Le-Roy-De-Bonneville:2021} and \citet{Zamansky:2022f}.

The volume forcing of the liquid phase in \eqref{eq:NS_big_bubbles} is given by
\begin{equation}
  	\bm{f}(\bm{x},t) = - \sum_{b=1}^{N_b} \bm{F}_{f\rightarrow b}(t) G_\sigma(\bm{x}-\bm{x}_b(t))\,,
	\label{eq:mom_exchange}
  \end{equation} 
where $\bm{F}_{f\rightarrow b}$ is the momentum exchange rate between the fluid and the bubble $b$, and $N_b$ is the number of bubbles.
$G_\sigma$ is the Gaussian kernel of the projection and $\sigma$ is its characteristic size which is of the order of the diameter of the bubble.
The latter is thus much larger than the mesh size: $\sigma \approx d > \Delta x$. 
Indeed, although the details close to the bubbles are filtered, the flow presents \textit{a priori} scales much smaller than the bubble size. 
These small-scale fluctuations result from the evolution of the turbulent wakes and their interactions.

The trajectory of the bubbles is obtained by solving the Newton's equation for each bubble. 
It involves the hydrodynamic force which depends on the velocity of the liquid (and its derivatives). 
\citet{Le-Roy-De-Bonneville:2021} considered that the bubble is subject to the drag force, the added-mass force, the inertia force of the fluid, as well as the buoyancy.
We have not retained the history force because it has a priori a negligible effect for large Reynolds numbers. 
On the other hand, when the velocity gradient is large at the scale of the bubble, the lift forces can certainly play a role.
Similarly, the anisotropic effects of drag and added mass related to a non-spherical bubble are also important.
We aim to reproduce the experiments of \citet{Riboux:2010} for millimetric air bubbles in water.
Given the Reynolds number of the bubbles and the Morton number, the bubbles clearly adopt a non-spherical shape \citep{Maxworthy:1996}.
However, to simplify the modeling of the problem, we consider that the bubbles are spherical, assuming that in the case of the homogeneous swarm, the anisotropic aspects are not essential.
Consistently, we have as well not retained the lift force. Note that the value of the lift coefficient, and even its sign, being very dependent on the shape of the object, it would be delicate to choose its value anyway.
Finally, considering that the density of the gas is very low we obtain for the dynamics of the bubble: 
\begin{equation}
C_M \dfrac{d \bm{v}_b}{d t} = \dfrac{3C_D}{4 d} (\bm{v}_b - \tilde{\bm{u}}_{f,b} )|\bm{v}_b - \tilde{\bm{u}}_{f,b}| 
	+(1+C_M) \dfrac{D \tilde{\bm{u}}_{f,b}}{D t} - \bm{g}   + \bm{F}_{\text{I},b}\,.
	\label{eq:force_big_bubble}	
\end{equation}
The drag coefficient is chosen, in agreement with the experiment of an isolated bubble, at $C_D = 0.35$ and the added-mass coefficient at $C_M=0.5$ in coherence with the spherical bubble hypothesis.
In this equation, the bubble force is calculated from the corrected liquid velocity $\tilde{\bm{u}}_{f,b}$, as proposed in \citet{Le-Roy-De-Bonneville:2021}, and briefly explained below. 
Note that  interactions between bubbles  are accounted through the liquid disturbances generated by the bubbles, which is well suited for the study of a homogeneous bubbly flows, but should probably be improved by adjusting of the drag and added-mass coefficients according to the local bubble distribution, in the spirit of the method proposed by \cite{Akiki:2017}.

Finally, the term $\bm{F}_{\text{I},b}$ is a repulsive force between bubbles. 
It is introduced to prevent the bubbles from overlapping and to ensure that the distance between bubbles remains greater than the characteristic size $\sigma$ of the momentum source.
The force depends on the distance between each pair of bubbles $r_{b,b'}= | \bm{x}_b-\bm{x}_{b'}|$ and is given by $\bm{F}_{\text{I},b} = \sum_{b'\neq b } -C_I \dfrac{\bm{x}_b-\bm{x}_{b'}}{r_{b,b'}}\exp(- r_{b,b'}^2/2r_I^2)$.
In the numerical simulation presented in the paper, we use the same value as prescribed in \citet{Le-Roy-De-Bonneville:2021}.
It has been verified that the modification of the values of these parameters does not modify significantly the simulations. 

The momentum exchanged between the bubble and the fluid via the volume force $\bm{f}$ in \eqref{eq:mom_exchange} is given by the sum of the drag force and the added-mass force, the contributions of the Tchen and Archimedes forces being already taken into account in the pressure term in a way consistent with the zero divergence of the flow \citep{Climent:1999,Le-Roy-De-Bonneville:2021}.

The fluid velocity, corrected for the influence of bubble $b$, is defined by introducing the perturbation due to the bubble:
\begin{equation}
	\tilde{\bm{u}}_{f,b}(x,t)= \bm{u}_{f}-\bm{u}^*_{f,b} \;.
	\label{eq:def_unperturbed_vel}
\end{equation}

Because of the non-linearity of the system, this immediately raises the question of the definition of the perturbation $\bm{u}^*_{f,b}$.
It is indeed not trivial to isolate the influence of a bubble among the fluctuations of the flow which include the effect of all the other bubbles. 
We propose here to define the perturbed field $\bm{u}^*_{f,b}$ as the flow generated by an isolated fictitious bubble, in a liquid at rest, which would have followed the same trajectory and exchanged as much momentum with the liquid phase as the actual bubble $b$. 
\citet{Le-Roy-De-Bonneville:2021} proposed an integral model to calculate $\bm{u}^*_{f,b}$, and its derivatives, valid for the case of bubbles at large Reynolds number.

In this model, the main assumptions to obtain $\bm{u}^*_{f,b}$ are that in the vicinity of the bubble (i) given the importance of the Reynolds number, the viscous term is neglected and (ii) the flow is considered quasi-parallel.
The details of the derivation can be found in \citet{Le-Roy-De-Bonneville:2021}, but after a few steps we obtain the following integral expression for $\bm{u}^*_{f,b}$.
\begin{equation}
	\bm{u}^*_{f,b}{\scriptstyle(\bm{x},t)}  = 
	\dfrac{1}{\rho_f} 
	\int_{0}^t \bm{F}_{f\rightarrow b}{\scriptstyle(s)} G_{\sigma}{\scriptstyle(\bm{x}-\bm{x}_b(s) + \bm{\ell}_{adv}(t,s) )}ds  \ ,
	\label{eq:u_pertu}
\end{equation}
where $\bm{\ell}_{adv}(t,s) = \int_s^t \tilde{\bm{u}}_{f,b} {\scriptstyle(\bm{x}=\bm{x}_b(s^{\prime}),s^{\prime})} ds^{\prime}$.
The velocity perturbation at a given position and time is obtained by integrating, over all previous instants, the momentum supplied by the bubble at the material point of the liquid at that specific position. 
The material point corresponding to the injection of momentum at an instant $s$ can be advected by the undisturbed flow, and will be found at a distance $\bm{\ell}_{adv}(t,s)$ at the instant $t>s$.
This $\bm{\ell}_{adv}$ term is essential to guarantee the Galilean invariance of the model.

We do not detail here the discretization of \eqref{eq:u_pertu}. 
The details of its numerical implementation can be found in \citet{Le-Roy-De-Bonneville:2021}. 
We just mention that we have developed an efficient algorithm to compute the history integral \citep[see also][]{Zamansky:2022f}, so the extra cost of computing the correction term is negligible.

It is interesting to note that, although the coupling between the two phases conserves momentum, it does not conserve energy \citep{Xu:2007b,Subramaniam:2014,Le-Roy-De-Bonneville:2021}.
Indeed, the power $P_b$ of the hydrodynamic force working at the bubble velocity is greater than the power $P_f$ of the diffuse force working at the fluid velocity: 
\begin{equation}
	P_b=\sum_b \bm{F}_{f\rightarrow b }\cdot \bm{v}_b  > P_f=\int dx^3 \bm{f}\cdot\bm{u}_f   \;.
\end{equation}
From a physical perspective, it is acceptable that energy is dissipated during the  coupling. We consider a coarse description of the continuous phase, in which the strong velocity gradients in the close vicinity of the bubbles are not described. 
The dissipation of kinetic energy into heat that occurs in the region surrounding the bubble at scales smaller than $\sigma$ cannot be calculated from the resolved velocity $\bm{u}_f$.
In the case of the isolated bubble, \cite{Le-Roy-De-Bonneville:2021} estimated analytically that 
\begin{equation}
	\dfrac{P_f}{P_b} = \dfrac{C_D }{64} \left( \dfrac{d_b}{\sigma}\right)^2.  
	\label{eq:power}
\end{equation}
Consequently, an important part of the mechanical energy is dissipated around the bubbles, in the boundary layer and the near wake.

 \subsection{Details of the simulations}

With this method we simulated the flow of a rising bubble swarm.
The parameters correspond to a 2.5-mm air bubble in water. 
The Reynolds number based on the terminal velocity of an isolated bubble is $Re_0=v_0 d / \nu = 760$.
A cubic domain of dimension $L/d = 70$ with tri-periodic boundary conditions is used. 
The characteristic size of the force projection kernel is $\sigma/d = 0.28$ and the resolution of the mesh is $\Delta x /d = 1/15$ which corresponds to 1024 points in each direction.
We can notice that the number of mesh points per bubble can seem important for a method which does not try to solve precisely the dynamics around the bubbles. 
However, one must keep in mind that (i) the resolution with interface tracking methods for such a Reynolds number of bubbles, requires about 100 meshes per bubble (or even more) to capture the boundary layer which develops on the bubble \citep{Du-Cluzeau:2019,Innocenti:2021} and (ii) the resolution is chosen here to capture the small scales which develop in the wakes, not too close to the bubbles, as we will see below.

We have simulated this flow for volume fractions $\alpha=1\%$, 2\%, 5\%, 7.5\% and 10\% corresponding to a number of bubbles ranging from $N_b=6500$ to 65000.
See the visualization of this flow for $\alpha=2\%$ and 10\% in figure \ref{fig:bubble_swarm_visu}.
In this figure, we can see the wakes generated by the passage of each bubble, and their interactions giving birth to the agitation induced by the rise of a bubble swarm.
Movies for the various volume fractions are also available in Supplementary Material.

Before examining the results, it is worth contextualizing the present work within the state of the art.
The main motivation for performing such simulations is to obtain a description of the fluctuations in the spectral domain, in order to get insights in the mechanisms controlling the peculiar dynamics of the bubble-induced turbulence. Experimental investigations, have shown some important features of the velocity spectrum (reviewed in the introduction), but further advances are facing severe limitations related to experimental constraints imposed by the presence of the bubbles. In this context, Direct Numerical Simulations (DNS) appear as a promising approach. However, they face two major problems. The first one is, of course, the limited resolution imposed by computational capabilities. The second one, less trivial, is the need of a consistent definition of the spectrum of a field quantities in the presence of jumps across interfaces between phases. Probably for those reasons, very few DNS studies have reported spectra in bubbly flows.

Two DNS studies dealing with a swarm of high-Reynolds number rising bubbles are worth mentioning: \cite{Roghair:2011} and \cite{Pandey:2020}. The resolution, the total number of bubbles and the volume fractions were $\Delta x /d = 1/20$, $N_b=16$, and $\alpha=5\%$ for a Reynolds number of the order of 1000 in \cite{Roghair:2011}, $\Delta x /d = 1/24$ and $N_b=40$, and $\alpha=1.7\%$ for $Re=546$ in \cite{Pandey:2020}. The resolution is only slightly better than ours, at the price of a much smaller number of bubbles and a limited volume fraction. Still, it is not fine enough to ensure that the smallest scales close to the bubbles are fully resolved. It can therefore not be concluded that they provide a better description of the bubble swarm than our present approach in which the approximation of the interfacial transfers across the interfaces is explicitly modeled.

Regarding the method of computing the spectrum, \cite{Roghair:2011} considered only intervals between bubbles where the velocity signal is continuous in order to mimic an experimental technique, which allowed them to compare with available experimental measurements. This approach avoids the problem of crossing interfaces, but suffers from the same limitations as experiments. In particular, it only gives access to the spectrum of the velocity.
 \cite{Pandey:2020} considered the entire flow field of the two-phase mixture and provided the (cumulated) spectrum of the terms of the energy budget.
 However, they did not address the problem of computing the spectrum of fields including discontinuities and Dirac delta functions. Moreover, they use an unusual definition of the terms of the energy budget involving the Fourier transform of the ratio between a Dirac  delta function and Heaviside function, which has not been proven to be valid \citep[see detail in][]{Ramirez:2023}. It is worth mentioning that in the single-fluid approach proposed here, the {\it a priori} filtering of the interfacial transfers by the coarse-grained method ensures that all computed fields are regular and that the spectral analysis does not suffer from any mathematical inconsistencies.
 
Despite their limitations, these two pioneering studies have produced interesting results. By comparing with previous point-bubble simulations, \cite{Roghair:2011} confirmed that the presence of wakes behind the bubble is necessary to obtain a $k^{-3}$ spectral subrange. \cite{Pandey:2020} suggested that the transfer between scales induced by inertia and interfacial forces play a role in the $k^{-3}$ spectral subrange. Nevertheless, due to their limitations, no quantitative comparisons can be made with the results discussed in the following section.

 \section{Comparison with experiments}\label{sec:exp}

\begin{figure}
   \centering
  \includegraphics[width=0.49\textwidth,height=0.35\textheight, keepaspectratio]{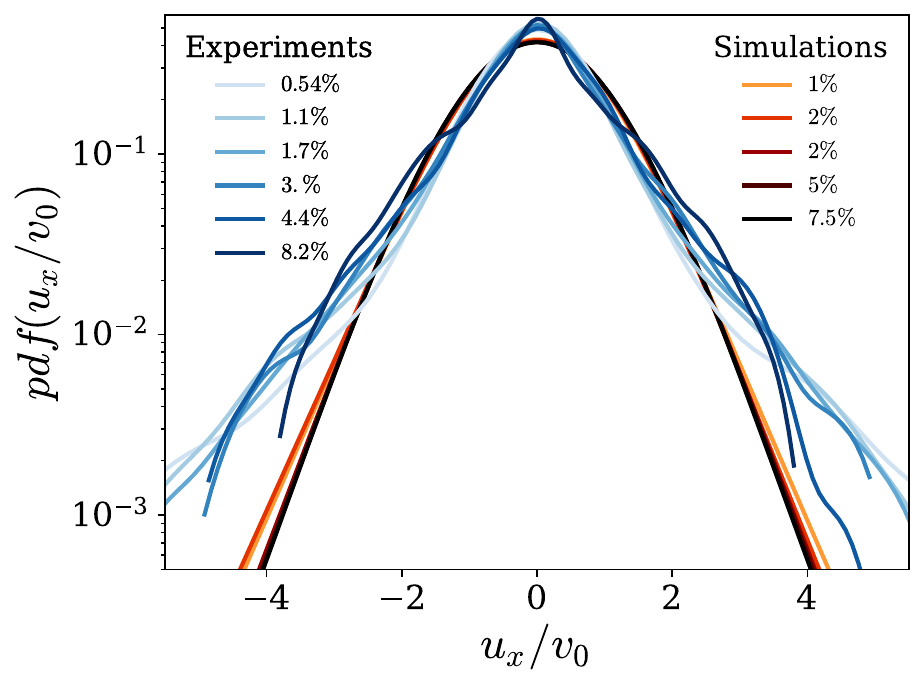}
   \includegraphics[width=0.49\textwidth,height=0.35\textheight, keepaspectratio]{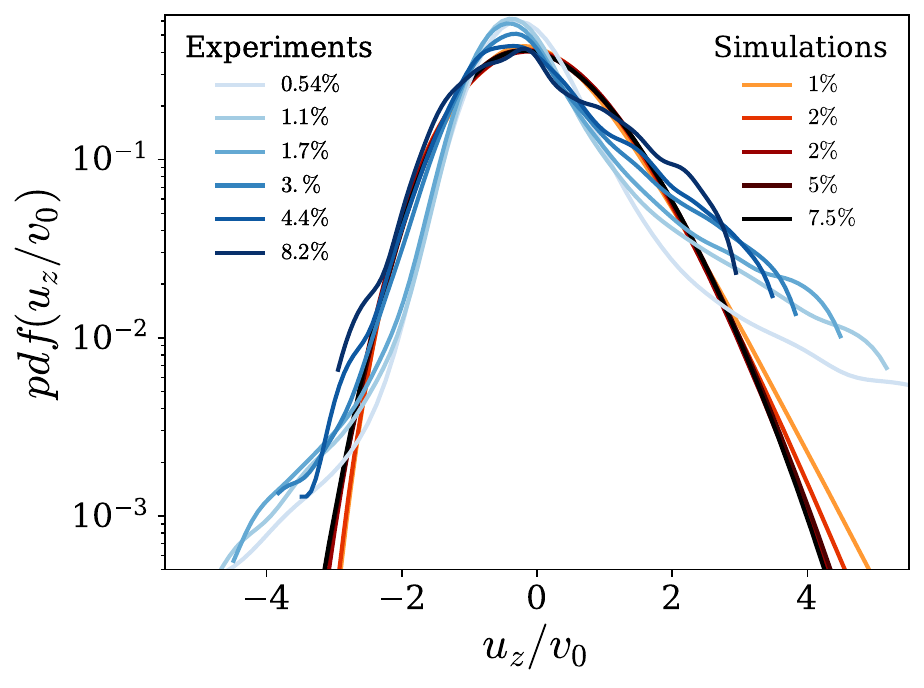} 
   \caption
   {PDF liquid velocity for the horizontal velocity component (left) and the vertical velocity component (right) from simulations at various $\alpha$ and comparison with the experiments of \citet{Riboux:2010} for $\alpha=0.54$, 1.1, 1.7, 3.0, 4.4 and $8.2\%$.
 	 }
 \label{fig:bubble_swarm_pdf_velocity}
 \end{figure}

Figure \ref{fig:bubble_swarm_pdf_velocity} shows the PDFs of the horizontal and vertical components of the liquid velocity obtained by the simulations for the different $\alpha$ and by the experiments of \citet{Riboux:2010}.
It is found that for both components the PDFs present an exponential decay and that the PDFs of the vertical component are clearly asymmetric.
We further observe that the normalized PDFs are nearly invariant with $\alpha$ as the central part of the experimental PDFs.
This behavior is the signature of the turbulence induced by the interactions between wakes.
For large fluctuations, the experimental PDFs show a second region characterized by a less steep exponential decay.
This behavior has been attributed to the large localized fluctuations very close to the bubbles \citep{Risso:2016}.
As this region is not described in our modeling, we indeed find that the second exponential part of the PDFs is not reproduced by the simulations.

For the same reason, the mean velocity of the bubbles decreases only very slightly with $\alpha$ according to the simulations, whereas experimentally, it is observed to decrease as $\langle v \rangle/v_0 \approx 0.6 \alpha^{-0.1}$ \citep{Riboux:2010}.
Also, the kinetic energies of the variances of the fluctuations of both the liquid and the bubbles are underestimated compared to the experiments.

\begin{figure}
   \centering
\includegraphics[width=0.49\textwidth,height=0.35\textheight, keepaspectratio]{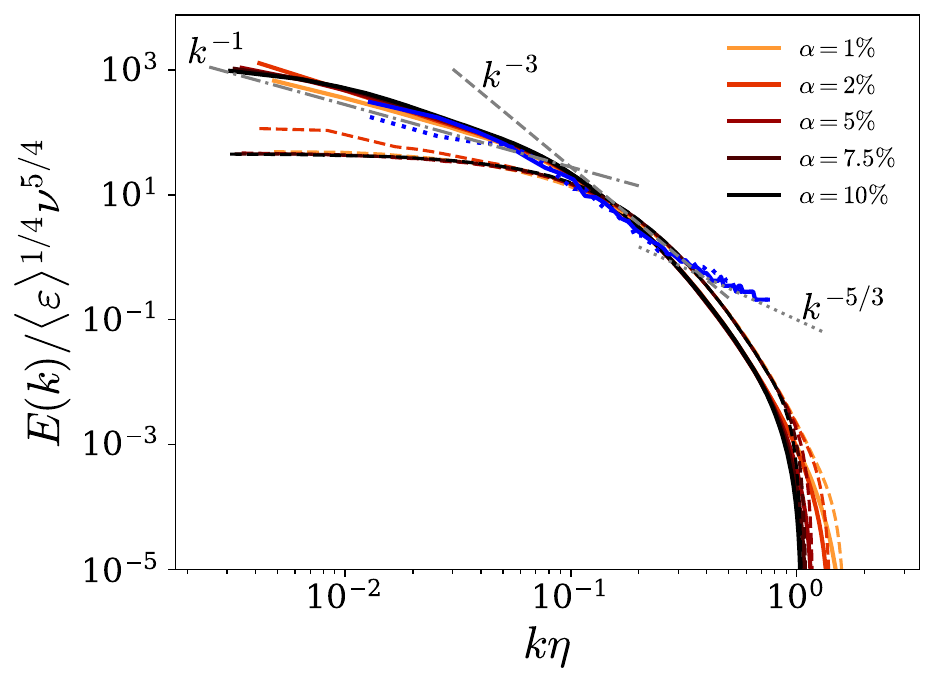}
\includegraphics[width=0.49\textwidth,height=0.35\textheight, keepaspectratio]{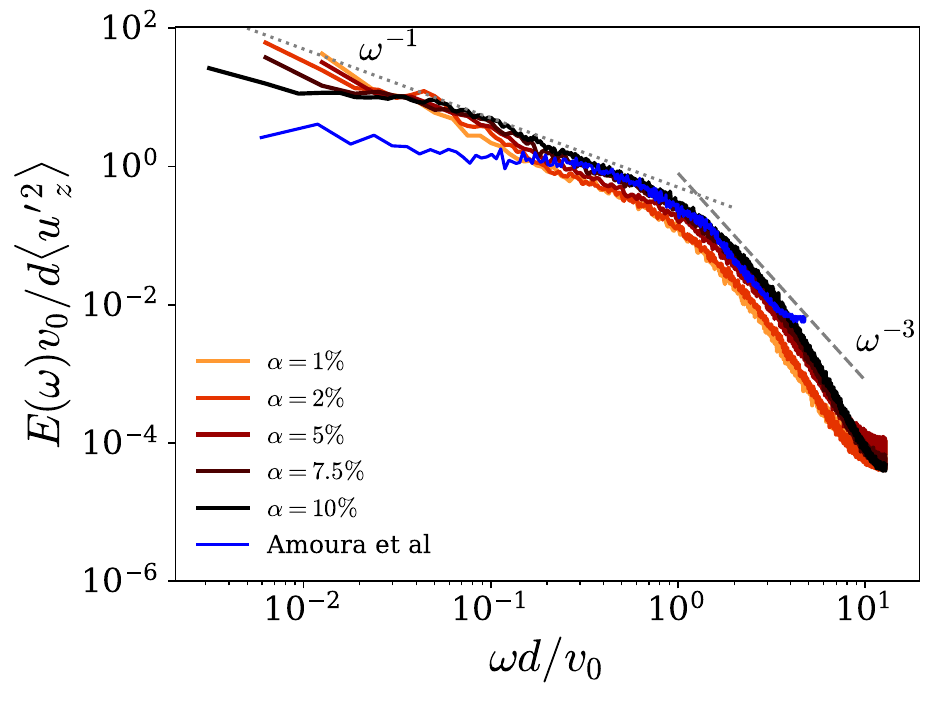}
   \caption
   {
   (Left)
   Longitudinal velocity spectra of the vertical component in the vertical direction ($E_z(k_z)$) (continuous lines) and of the horizontal component in the horizontal direction (dashed lines) from the simulations at various $\alpha$ ($E_x(k_x)$).
   Comparison, with the experiments of  \citet{Riboux:2010} for $\alpha=2.5\%$ and $d=2.5mm$ in blue, and with the power law $k^{-1}$ (gray dot-dashed line), $k^{-3}$ (gray dashed line) and $k^{-5/3}$ (gray doted line).\\
   (Right) Frequency spectra of the liquid velocity at the bubble position from the numerical simulation and comparison with the experimental spectra of the liquid velocity of the flow past a random array of fixed spheres \citep{Amoura:2017} in blue and with the $\omega^{-1}$ and $\omega^{-3}$ power laws.
 	 }
 \label{fig:bubble_swarm_spectra1D_velocity}
 \end{figure}

Figure \ref{fig:bubble_swarm_spectra1D_velocity} compares the longitudinal spectra\footnote{That is to say $E_x(k_x)=\int 1/2 \phi_{xx}(\bm{k}') \delta(\bm{k}'.e_x-k_x)d^3\bm{k}' $ and $E_z(k_z)=\int 1/2 \phi_{zz}(\bm{k}') \delta(\bm{k}'.e_z-k_z)d^3\bm{k}' $ with $\phi_{ij}(\bm{k}) = \sum_{\bm{k}'}\langle \hat{u}_{i}(\bm{k}')\hat{u}_{i}^*(\bm{k}') \rangle  \delta(\bm{k}-(\bm{k}')) $ and $\hat{\bm{u}}$ the coefficients of the Fourier series of the velocity field $ \bm{u}_f(\bm{x})= \sum_{\bm{k}}  e^{i\bm{k}.\bm{x}}\hat{\bm{u}}(\bm{k})$.}
 of the vertical and horizontal velocities $E_z(k_z)$ and $E_x(k_x)$ obtained experimentally and by simulations.
Note that, with this simulation approach, the spectra of the velocity field are very easy to obtain because the $\bm{u}_f$ field is smooth, whereas with a DNS type approach, as well as in experiments, the velocity of the liquid is not defined everywhere, which poses a number of problems for spectral analysis.
 Here the approximations are made prior to the simulation, at the modeling phase, and there is no particular precaution to take for the calculation of the spectra.
We can see that the spectra of the vertical and horizontal components are in fairly good agreement with the experiment.
 In particular, the simulations seem to reproduce a $k^{-3}$ evolution  of the spectra as experimentally observed  on small scales (large wavenumbers) and a $k^{-1}$ decay at large scales.

 However, we note, on the one hand, that the simulations underestimate the kinetic energy at small scales. We attribute this to the lack of near-bubble resolution, which leads to an underestimation of the power injected at the bubble scale (as discussed in the previous section).
 On the other hand, we also notice that the larger scales of the horizontal component are also underestimated, due to the absence of bubble trajectory oscillations, which are expected to enhance the redistribution of the fluctuating energy between the vertical and horizontal components.

It has also been reported that experimentally the spectra are invariant with the volume fraction and the bubble diameter.
When the spectra of the numerical simulations are normalized by the injected power and by the viscosity, we observe the same invariance of the spectra of the numerical simulation with $\alpha$.

In figure \ref{fig:bubble_swarm_spectra1D_velocity}, we present the frequency spectra of the vertical liquid velocity measured at the position of the bubbles (the true velocity, not the corrected one). 
These spectra are compared with the frequency spectrum of the velocity of the flow passing through an array of spheres held at a fixed position obtained  experimentally by \citet{Amoura:2017} for a Reynolds number, based on the sphere diameter, of 600. 
Although the two flows are different, in both cases these spectra can be considered as characterizing the fluctuations of the liquid in the frame of reference moving with the bubbles.
It can be seen that the simulations and the experiment show again a remarkable agreement.
For frequencies lower than $d/v_0$ the spectrum shows a $\omega^{-1}$ behavior, while at high frequencies, the cutoff is much stronger with a slope close to $\omega^{-3}$. 
It is interesting to note that this behavior seems invariant with $\alpha$ in the simulations, while experiments have reported that it is invariant with the Reynolds number, provided that $Re>200$ \citep{Amoura:2017}.

To summarize, while the bubble rising velocity and the total energy production is underestimated because of the filtering of the energy transferred from the bubble to the fluid, the interactions between wakes are well reproduced. Since this essential physical mechanism is correctly accounted for, the normalized spectra of the velocity are representative of real flows.

\section{Characteristic scales}\label{sec:scales}

\begin{figure}
   \centering
   \includegraphics[width=0.49\textwidth,height=0.35\textheight, keepaspectratio]{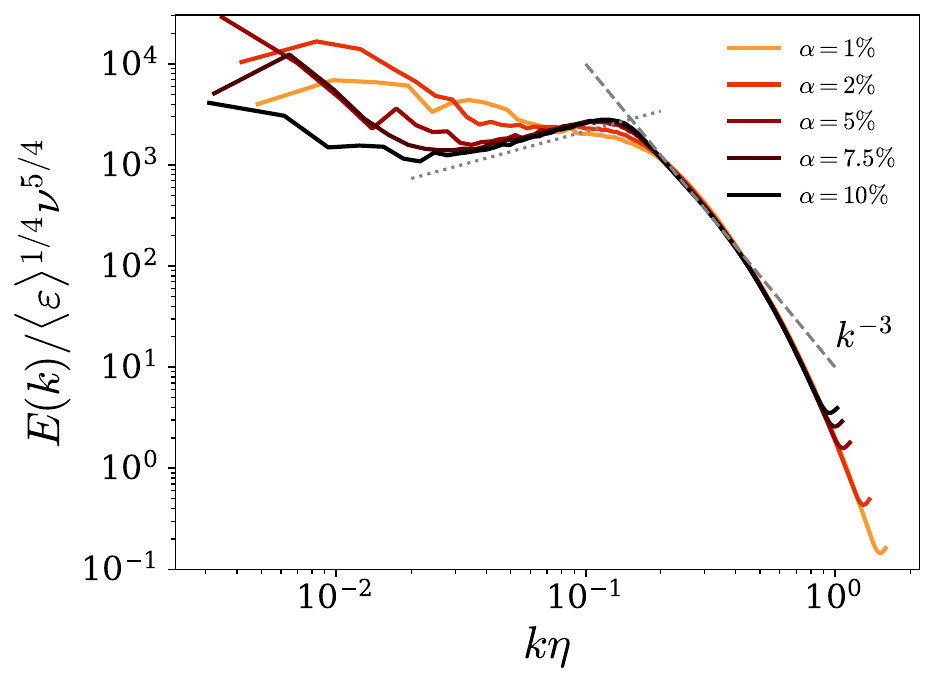}
   \caption
   {Spherically averaged  spectra of the velocity  field $E$ from the numerical simulation at various $\alpha$. 
   The $k^{-3}$ power law is shown by a gray dashed line.  (The gray dotted line corresponding to a $k^{2/3}$ power law is just here as a guide for the eyes). }
 \label{fig:bubble_swarm_spec}
 \end{figure}

The spherically averaged spectra of the velocity\footnote{\textit{i.e.}, $E(k)=\int 1/2 \phi_{ii}(\bm{k}') \delta(|\bm{k}'|-k)d^3\bm{k}' $. \label{note:sphere_spec}} are shown in figure \ref{fig:bubble_swarm_spec}.  
Contrary to the longitudinal spectra presented in figure \ref{fig:bubble_swarm_spectra1D_velocity}, the three-dimensional spectra show a more complicated evolution with $k$ as well as a rather clear dependence with $\alpha$ at large scales. 
Several regions can be distinguished. 
The local maximum, located around $k \eta \approx 2$ on the figure, coincides, as we will see, with the scale of the bubbles which gives the cut-off scale of the energy injection. 
We see that for larger wavenumbers, a region in $k^{-3}$ clearly develops as $\alpha$ increases.
The local minimum, located at large scales, corresponds to the wake scale. 
Between these two scales, the energy spectrum depends on $\alpha$ and corresponds to the scales directly influenced by the wakes.
The fact that the spherically averaged spectra $E$  show such a qualitative difference at large scales with the spectra averaged over planes obviously indicates that the flow has a strong anisotropy at these scales. We will come back to the characterization of the anisotropy below.

The flow being presumably dominated by the wakes of the bubbles and their interactions, an essential scale of the flow is the characteristic length of the wakes.
To determine the latter, we consider the mean field conditioned on the position of a bubble (equivalent to a spatial phase average): 
\begin{equation}
	\langle \bm{u}_f \rangle_b(\bm{x}) = \dfrac{1}{T}\int dt \dfrac{1}{N_b}\sum_{b=1}^{N_b} \bm{u}_f(\bm{x}-\bm{x}_b(t),t) \ .
\label{EQ5}
\end{equation}

\begin{figure}
   \centering
   \includegraphics[width=0.25\textwidth,height=0.45\textheight, keepaspectratio]{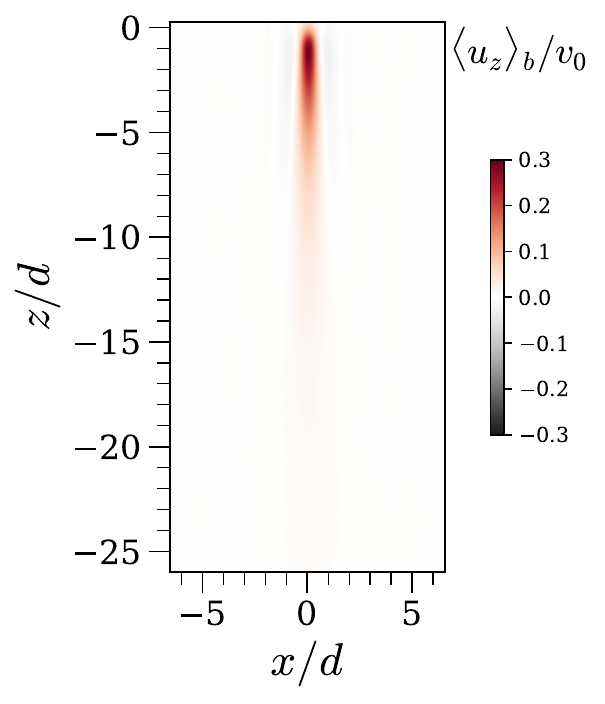}
    \includegraphics[width=0.4\textwidth,height=0.35\textheight, keepaspectratio]{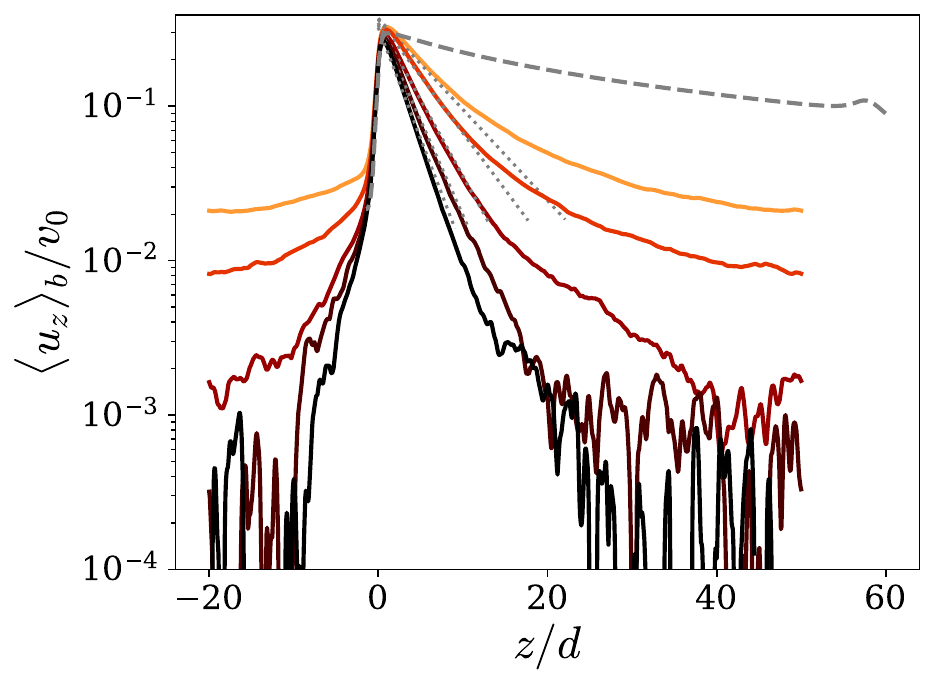}
	\includegraphics[width=0.4\textwidth,height=0.35\textheight, keepaspectratio]{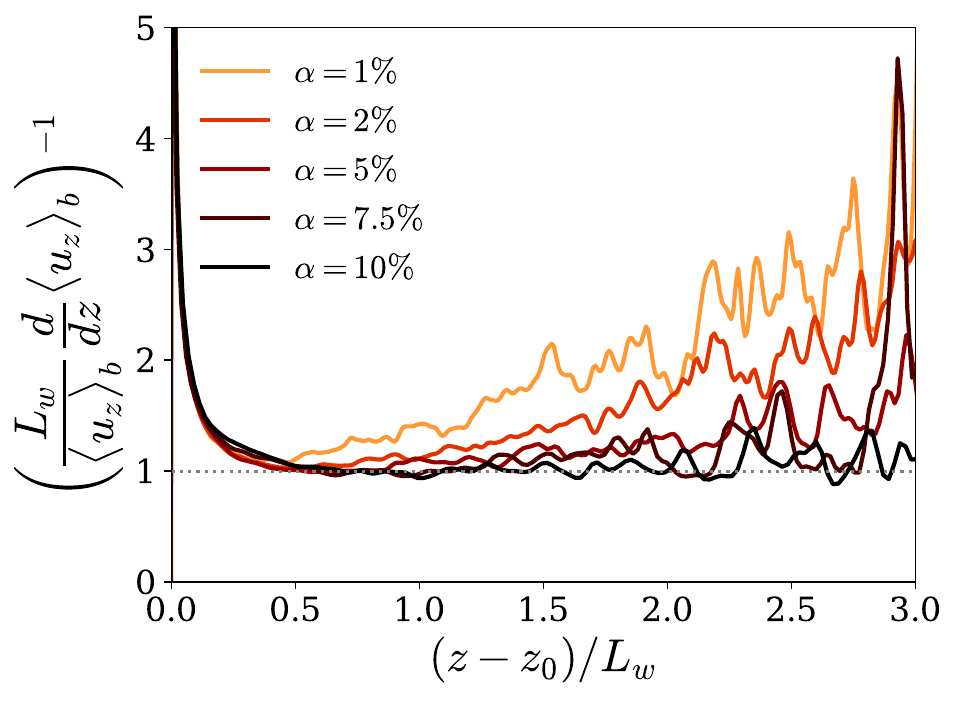}
   \caption
   {
	   Upper left: cross-section of the vertical velocity conditionally averaged to the bubble position (eq.~ \ref{EQ5}): $\langle u_{f,z} \rangle_b$ for $\alpha=5\%$.
	   Upper right: evolution of the $\langle u_{f,z} \rangle_b$ along the vertical axis passing through the origin for the various $\alpha$ and for an isolated bubble (in gray dashed line), comparison with the exponential decay with a rate $L_w(\alpha)$ (in gray dotted line). 
	   For ease of reading, we take advantage of the periodicity of the domain, and move the increassing part of $\langle u_{f,z} \rangle_b(z)$ upstream the bubble.
	   Lower: inverse of the logarithmic derivative, $\dfrac{1}{\langle u_z \rangle_b }\dfrac{d \langle u_z \rangle_b }{d z}$, of the former quantity,  normalized by the characteristic wake length $ L_w$. For this figure the vertical position is shifted by $z_0$ corresponding to the vertical position of the maximum of $\langle u_{f,z} \rangle_b$ .
	 }
 \label{fig:bubble_swarm_mean_wake}
 \end{figure}

This mean field is illustrated in figure \ref{fig:bubble_swarm_mean_wake} for the case $\alpha=5\%$. 
This figure also shows the evolution of the mean vertical velocity along the vertical axis passing through the bubble for the different volume fractions, as well as for an isolated bubble. 
The first observation is that the wakes are much shorter in the case of the bubble swarm than in the case of an isolated bubble. 
By plotting the logarithmic derivative of the wakes, also shown on figure \ref{fig:bubble_swarm_mean_wake}, we see that the wakes present a self-similar evolution for all $\alpha$ and that the mean velocity decreases exponentially with $z$.
This remarkable feature is in agreement with the experimental results presented by \citet{Risso:2008}.
This exponential decay of the wakes is likely due to the canceling of the vorticity between neighboring wakes, as proposed by \citet{Hunt:2002}.
 
\begin{figure}
   \centering
   \includegraphics[width=0.49\textwidth,height=0.35\textheight, keepaspectratio]{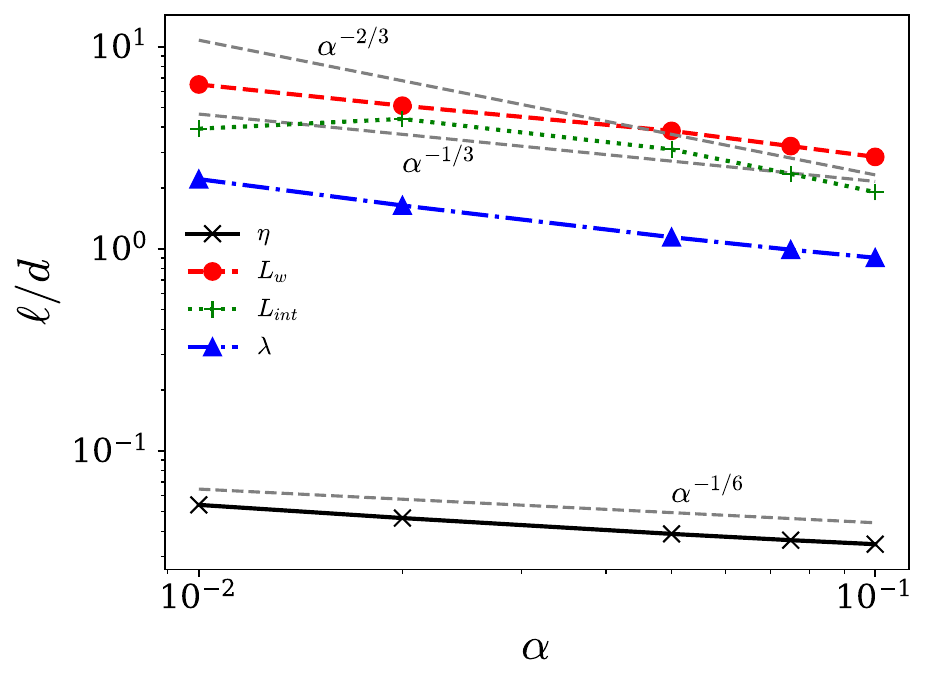}
   \caption
   {Evolution of the characteristic scales $L_w$, $\lambda$, $\eta$ and $L_{int}$ in the simulations of the bubble swarm with $\alpha$.
   Comparison with the power law $\alpha^{-1/3}$ in dashed lines, $\alpha^{-1/6}$ in dotted lines and $\alpha^{-2/3}$ in dot-dashed lines.
 	 }
 \label{fig:bubble_swarm_scales}
 \end{figure}

We  choose the relaxation length of the exponential as the characteristic scale of the wakes $L_w$.
The evolution of the ratio $L_w/d$ as a function of $\alpha$ is presented in figure \ref{fig:bubble_swarm_scales}. 
We can see that the length of the wakes shows evolution in $L_w \sim d \, \alpha^{-1/3}$. 
One can interpret this evolution as a simple geometrical relation, considering that the wakes tend to screen each other.  
It should be noted, however, that this is quite a notable difference compared to the experiments of \citet{Risso:2008} where the characteristic length of the wakes was observed to be independent of $\alpha$.
This certainly reflects that there is an additional dependence of $L_w$ on $C_D$, since in the experiments the average speed of the bubbles decreases with $\alpha$.

From this characteristic wake length, we define an inverse timescale $f=v_0/L_w$.
This frequency $f$ can be considered as imposing a shear-rate scale to small scales $k\gtrsim 1/d $.
This assumption allows us to estimate the average dissipation rate in the simulations as 
\begin{equation}
	\langle \varepsilon \rangle = \nu f^2 .
	\label{eq:dissip_bubble_swarm}
\end{equation}
Equivalently, we can interpret $L_w$ as a Taylor length scale based on the velocity $v_0$, $\lambda= \sqrt{\nu v_0^2/\langle \varepsilon \rangle}$.
This is confirmed in figure \ref{fig:bubble_swarm_scales} which shows that the evolution of $\lambda/d$ varies as $L_w/d$ in $\alpha^{-1/3}$.

The volume averaged  power injected in the system corresponds to $\langle P_{tot} \rangle =n_b\langle P_b\rangle$ with $n_b$ the average number of bubbles per unit of volume and $P_b=\bm{F}_{f\rightarrow b}.\bm{v}_b$ the power supplied by bubble $b$. 
In the steady regime, this quantity is approximately given by $\langle P_{tot} \rangle = \alpha g v_0$ and as we have discussed above, it is larger than the power effectively received by the fluid in our simulations: $\langle P_{tot} \rangle$ > $\langle P_f \rangle = \langle \varepsilon \rangle$. 
We will thus interpret $\langle P_f \rangle$ as the mechanical energy effectively injected in the wakes. 
Combining the previous relations, we find $\langle P_{tot} \rangle/ \langle \varepsilon \rangle \sim \alpha Re_0 C_D (L_w/d)^2$ with $C_D = 4 gd/3v_0^2$. 
Therefore at $C_D$ and $Re_0$ constant, the proportion of energy injected in the wakes decreases as $\alpha^{-1/3}$.

From the estimate \eqref{eq:dissip_bubble_swarm} of the mean dissipation rate, we compute the dissipative scale $\eta = \nu^{3/4} \langle \varepsilon \rangle^{-1/4}= \sqrt{\nu/f} \sim d\,  Re_0^{-1/2}\alpha^{-1/6}$. 
It is this scale that is used to normalize the spectra presented in figures \ref{fig:bubble_swarm_spectra1D_velocity} and \ref{fig:bubble_swarm_spec}.

For $\alpha\geq 5\%$ we observed (not shown here, for brevity) that the kinetic energy of the liquid is invariant with $\alpha$ and is commensurate with $v_0^2$.
 Consequently we estimate the integral scale $L_{int} = \langle K\rangle^{3/2}/ \langle \varepsilon \rangle$ to vary as $L_{int} \sim d \, Re_0 \alpha^{-2/3}$. This behavior is observed for $\alpha\geq 5\%$ in figure \ref{fig:bubble_swarm_scales}.
Note that, in the experiments of \citet{Riboux:2010}, the liquid kinetic energy was observed to vary roughly as $\alpha v_0^2$.
The discrepancy of the numerical simulations with the experiments is once again attributed to the absence of the fluctuations localized in the vicinity of the bubbles, which scales with $\alpha$.

\section{Spectral analysis of the bubble-induced turbulence}\label{sec:spec}
  
\begin{figure}
   \centering
	\includegraphics[width=0.49\textwidth,height=0.35\textheight, keepaspectratio]{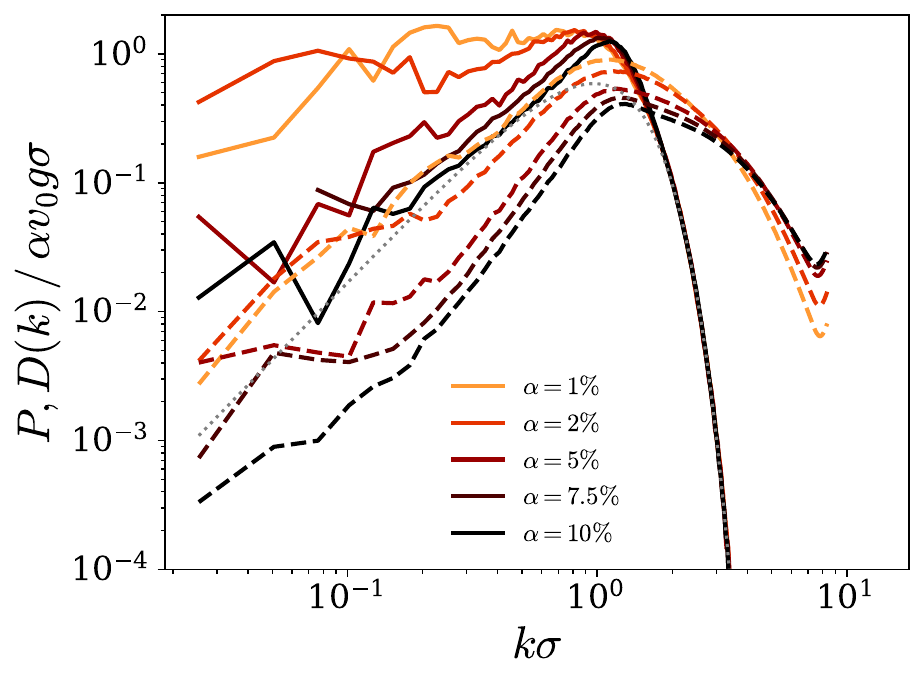}
    \includegraphics[width=0.49\textwidth,height=0.35\textheight, keepaspectratio]{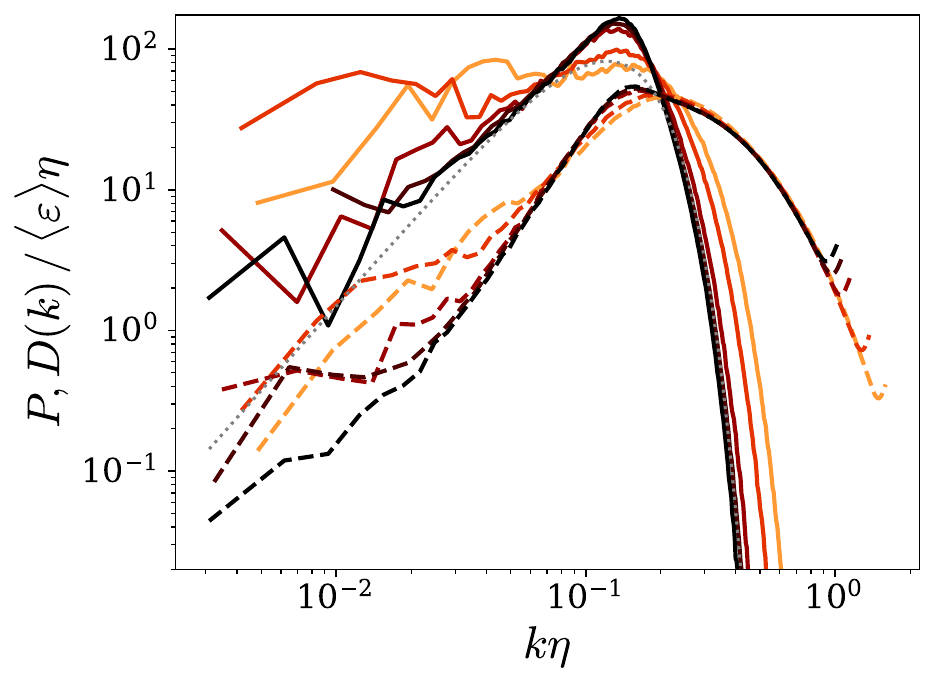}
   \caption
   { Evolution of $P$ in continuous line and of $-D$ in dashed lines for the various $\alpha$. 
   Comparison with $  \alpha g v_0 \sqrt{2/\pi} (k\sigma)^2\exp(-k^2\sigma^2)$ in dotted line.
   Left: the power density and wavenumber are normalized by  $\alpha g v_0d$ and $\sigma$ respectively.
   Right: the power density and wavenumber are normalized by the dissipative scales $\varepsilon \eta $ and $\eta$ respectively.
 	 }
 \label{fig:bubble_budg}
 \end{figure}

In order to identify the different regions of the spectra and to explain the observed scaling laws, we are interested in the spectral decomposition of the energy balance: 
\begin{equation}
\displaystyle
\frac{d }{dt}  E(k)  = T(k) -D(k)  + P(k) .
\label{eq:NS_spec_budg}	
\end{equation}
The terms of the right-hand side correspond respectively to the inter-scale energy transfer from a scale $k$  ($T$),  the kinetic energy dissipation at a scale $k$ ($D$) and the rate of energy injected by the bubbles ($P$). 
The expressions of these different terms are obtained from the Navier-Stokes equation \eqref{eq:NS_big_bubbles}.
The transfer term $T$ is the contribution of the non-linear terms: 
$T(k) =  \int \sum_{k'} \left[ -i k'_j ( \widehat{u_iu_j}\widehat{u}_i^*)\right] \delta(\bm{k'}-\bm{k})\delta(|\bm{k}|-k) d^3\bm{k} + C.C.$, where $+ C.C.$ denotes the complex conjugate terms,
 $D(k)= 2 \nu \mathrm{k}^2 E(k)$, 
$P(k) $ is the integral over the wave numbers $|\bm{k}|=k$ of the real part of $\hat{f}_i \hat{u}_i^*$, and  $\widehat{.}$  denotes the Fourier transform.
At steady state, the left-hand term of \eqref{eq:NS_spec_budg} is zero, so $T= D-P$.
We present in figure \ref{fig:bubble_budg} the terms $P(k)$ and $D(k)$ for the various $\alpha$.
We can see that the production term presents a cut-off for $k>1/\sigma$ (we recall that $\sigma/d = 0.28$), and that on large scales it grows as $k^2$ for the largest $\alpha$, while it is roughly constant for small $\alpha$.
Concerning the dissipation term, we notice that it also presents a peak around $k\sim 1/\sigma$. 
At large scales, the production dominates compared to the dissipation, which implies that $P(k) \approx -T(k)$.
On the other hand, the dissipation dominates on small scales which means that $D(k)\approx T(k)$.
The absence of scale separation between production and dissipation peaks means that this flow does not present an inertial zone.
These budgets also show that there is no range of scales in which there is an equilibrium between $P$ and $D$.
This contradicts the hypothesis made by \citet{Lance:1991} to explain the presence of a $k^{-3}$ zone in the velocity spectra.
Furthermore, we notice that the $k^{-1}$ region of $D$, which corresponds to the $k^{-3}$ region of the velocity spectra, is observed in the crossover between the production dominated scales and the dissipation dominated scales.

 \begin{figure}
   \centering
   \includegraphics[width=0.49\textwidth,height=0.35\textheight, keepaspectratio]{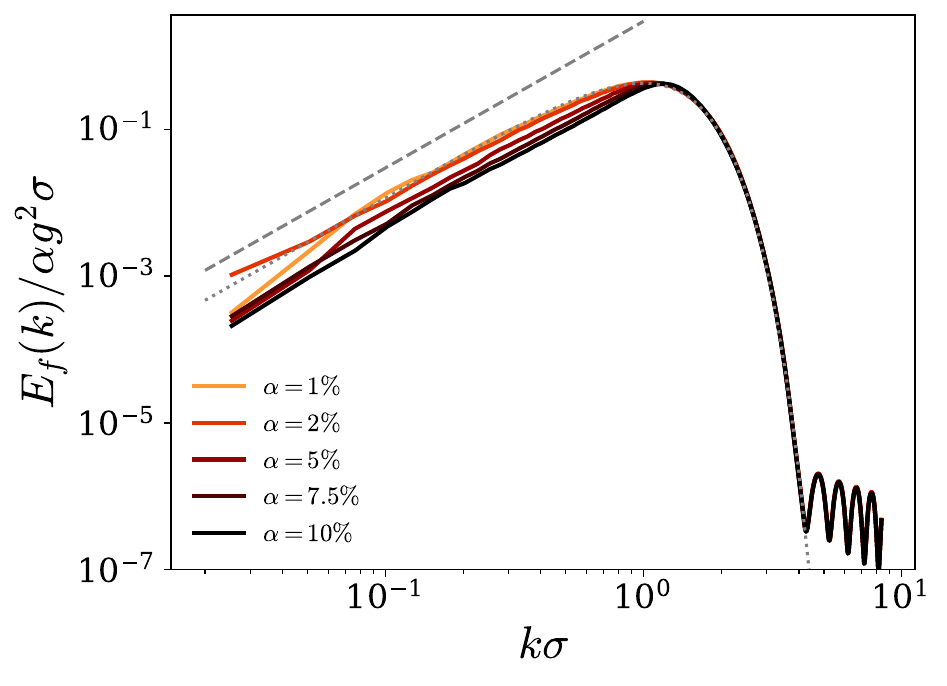}
   \caption
   { Spectra $E_f$ of the force at the various $\alpha$ and comparisons with \eqref{eq:spec_f} in dotted line and with the power law $k^2$ in dashed line.}
 \label{fig:Spect_forc_bubble}
 \end{figure}

To interpret the behavior of the production term $P$, we study the spectrum of the force applied to the flow, $E_f(k)$, corresponding to the spherical integration of $\hat{f}_i\hat{f}_i^*$.
From the expression of the coupling force between the phases \eqref{eq:mom_exchange}, we can  obtain the following analytical expression for $E_f$: 
 \begin{equation}
 	E_f(k) / \alpha \sigma g^2  =  \frac{1}{12\pi}(d/\sigma)^3 (k\sigma)^2e^{- (k\sigma)^2} .
	\label{eq:spec_f}
 \end{equation}
This expression is obtained by assuming that (i) the positions of the bubbles are independent from each other and that (ii) the fluctuations of the rate of momentum exchanged between the bubble and the liquid are small: $\langle \bm{F}_{f\rightarrow b}^2 \rangle \approx \langle \bm{F}_{f\rightarrow b} \rangle^2 = (\rho g \pi d^3/6 )^2 $.
The spectra of the force is presented in figure \ref{fig:Spect_forc_bubble}. 
We note that, at all volume fractions, the agreement with the proposed expression is relatively good.
We distinguish two regions: a region which grows as $k^2$ which reflects the equipartition of the fluctuations of the forces at large scales ($k<1/\sigma$) and an exponential decrease imposed by the Gaussian projection kernel $G_\sigma$ for $k>1/\sigma$.
Note that the oscillations observed at the end of the spectra are due to the sharp cut-off of the kernel $G_\sigma(\bm{x}-\bm{x_b})$ for $|\bm{x}-\bm{x_b}|> 3 \sigma$.
We notice that for high volume fractions, the positions of the bubbles are not really independent anymore, because they cannot overlap, which explains that the prefactor of the $k^2$ increase at small $k$ is reduced compared to \eqref{eq:spec_f}.

It is more difficult to propose an analytical estimate of the spectrum $P$ of  the work of $ \bm{F}_{f\rightarrow b}$. However, the spectrum $E$ of $u$ is much less steep than the spectrum $E_f$ of $\bm{F}_{f\rightarrow b}$. 
At large scales, $E$ is rather flat compared to the $k^2$ evolution of $E_f$ and, at small scales, $E$ shows a power-law decay compared to the exponential cutoff $E_f$. It is thus reasonable to expect $P$ to behave similarly to $E_f$. Considering that $P$ is dominated by buoyancy, we obtain the following estimate:
\begin{equation}
	P/ \alpha g v_0 \sigma \sim (k\sigma)^2 e^{- (k\sigma)^2}
\end{equation}
Indeed $P$ presents an exponential damping for $k>1/\sigma$, that overlaps for all $\alpha$ when normalized by $ \alpha g v_0 \sigma $ as can be seen in  figure \ref{fig:bubble_budg} (left).
For $k<1/\sigma$ and $\alpha>2\%$, $P$ increases roughly as $k^{2}$ in agreement with the previous relation.
The underestimation of $P$ at large scales is attributed to the the large fluctuations of the injected power $\langle P_f^2 \rangle \gg \langle P_f \rangle^2$.
For small $\alpha$, $P$ no longer follows the proposed relation on large scales  due to the presence of large structures in the flow leading to a significant correlation of the liquid velocity between distant bubbles.

For completeness, we present in figure  \ref{fig:bubble_budg} (right) the production and dissipation terms normalized by the dissipative scales $\langle \varepsilon \rangle $ and $\eta$ for the various $\alpha$. 
Consistently with the velocity spectra shown previously in figure \ref{fig:bubble_swarm_spec}, with this normalization, we observed that the values of $D$ of all  $\alpha$ overlap at high wavenumbers (typically $k>1/2\eta$).
From the estimates of the characteristic scales of the simulations presented in the previous section, we have $d/\eta = Re_0^{1/2}\alpha^{1/6}$, indicating that the gap between the production-dominated scales and the dissipation-dominated scales increases, slowly, with $\alpha$. 
It seems that the  $k^{-1}$ subrange of $D$ is observed in this gap of scales, provided that $\alpha$ is large enough.  

In conclusion, we consider that for $ k < 1/d$ the flow structure is driven by the interactions between wakes, while  in the range $1/d< k < 1/\eta$ the strong damping of the wakes imposes a shear scale.

This assumption of constant shear rate $f$ across scales allows us to explain the presence of a power law in $k^{-3}$ for the flow, thanks to a matching argument similar to that proposed by Kolmogorov in 1941.
It is assumed that at scales that are small compared to the length of the wakes ($k \gg 1 /L_w$) the structure of the flow depends only on the diameter of the bubbles $d$, the viscosity, and the shear rate $f$ : 
\begin{equation}
	E=E(k; d, \nu, f ).
\end{equation}
It is assumed that at scales larger than $\eta = \sqrt{\nu/f}$, one can neglect the effect of viscosity.
Therefore, in this limit we can write:
\begin{equation}
	E=E_I(k; d,f ) = d^3 f^2 \Phi_I(kd)
\end{equation}
where $\Phi_I$ is a dimensionless function. 
Conversely, at scales much smaller than the bubble size, we will suppose that the diameter does not play a role anymore, and we will make the hypothesis that 
\begin{equation}
	E=E_S(k; \nu, f) = \nu^{3/2} f^{1/2} \Phi_S(k\eta)
\end{equation}
where $\Phi_S$ is another dimensionless function. 
Finally, if we assume that for a range of intermediate scales ($1/d \ll k \ll 1/\eta$), the two previous relations remain valid, we have $E_S(k)=E_I(k)$.
Since $kd$ and $k\eta$ can vary independently, the previous equality can hold only if the following expressions are constant:
\begin{equation}
	(kd)^3 \Phi_I(kd) = (k\eta)^3 \Phi_S(k\eta) = c
\end{equation}
 This gives us for the velocity spectra:
 \begin{equation}
 	E(k)=c f^2 k^{-3}
	\label{eq:k-3}
 \end{equation} 
 in a range of scales where the shear rate can be considered constant.
  
 \bigskip
  
The temporal spectra of the velocity seen by the bubbles (presented in figure \ref{fig:bubble_swarm_spectra1D_velocity}, right) is influenced by the fact that the bubbles cut the wake of other bubbles. 
Thus the high frequencies of the temporal spectrum are dominated by the Doppler shift due to the high-speed crossing, of the order of $v_0$, of the dissipative structures of the flow.
So using \eqref{eq:k-3} and taking the argument of  \citet{Tennekes:1975}, with $\omega \sim v_0 k$, we can estimate the high-frequency behavior of the frequency spectra:
\begin{equation}
	E(\omega) = E(k) \dfrac{k}{\omega} = f^2v_0^2 \omega^{-3}\; .
\end{equation}
The temporal spectra from both the experiments and the simulations present  a $\omega^{-3}$ zone at high frequency.

 \bigskip
  
We have seen that at large scales ($k\ll 1/d$), where the flow is dominated by wake interactions, there is a balance between production and inertia.
At these scales we notice that the one-dimensional spectra of the velocity present a $k^{-1}$ dependence. 
This means that each decade contains an equal amount of energy.
This behavior can be explained by the intermittence of the wake passages, giving rise to an alternation between periods of activity and calm \citep{Mandelbrot:1999}.
At these scales, the characteristic velocity no longer depends on a specific length scale and corresponds to the typical velocity of the bubbles $v_0$.
These periods of activity (the wakes) are characterized by their self-similar character \citep{Bak:1987,Marinari:1983} and present a variable intensity and duration, whereas the calm periods follow a Poissonian distribution reflecting the quasi-uniform distribution of the bubbles. 
The absence of a characteristic length leads directly to the absence of a characteristic time for the fluctuations. 
Hence the frequency spectrum of the velocity also shows a decay close to $\omega^{-1}$ at low frequency.
As pointed out by \citet{Mandelbrot:1967}, these  behaviors in $\omega^{-1} $ and $ k^{-1}$ must also be connected to the non-Gaussianity of the velocity distributions as well as to a long-range correlation of the velocity.

\section{Characterization of the anisotropy}\label{sec:aniso}

\begin{figure}
     \centering
	 \includegraphics[width=0.25\textwidth,height=0.35\textheight, keepaspectratio]{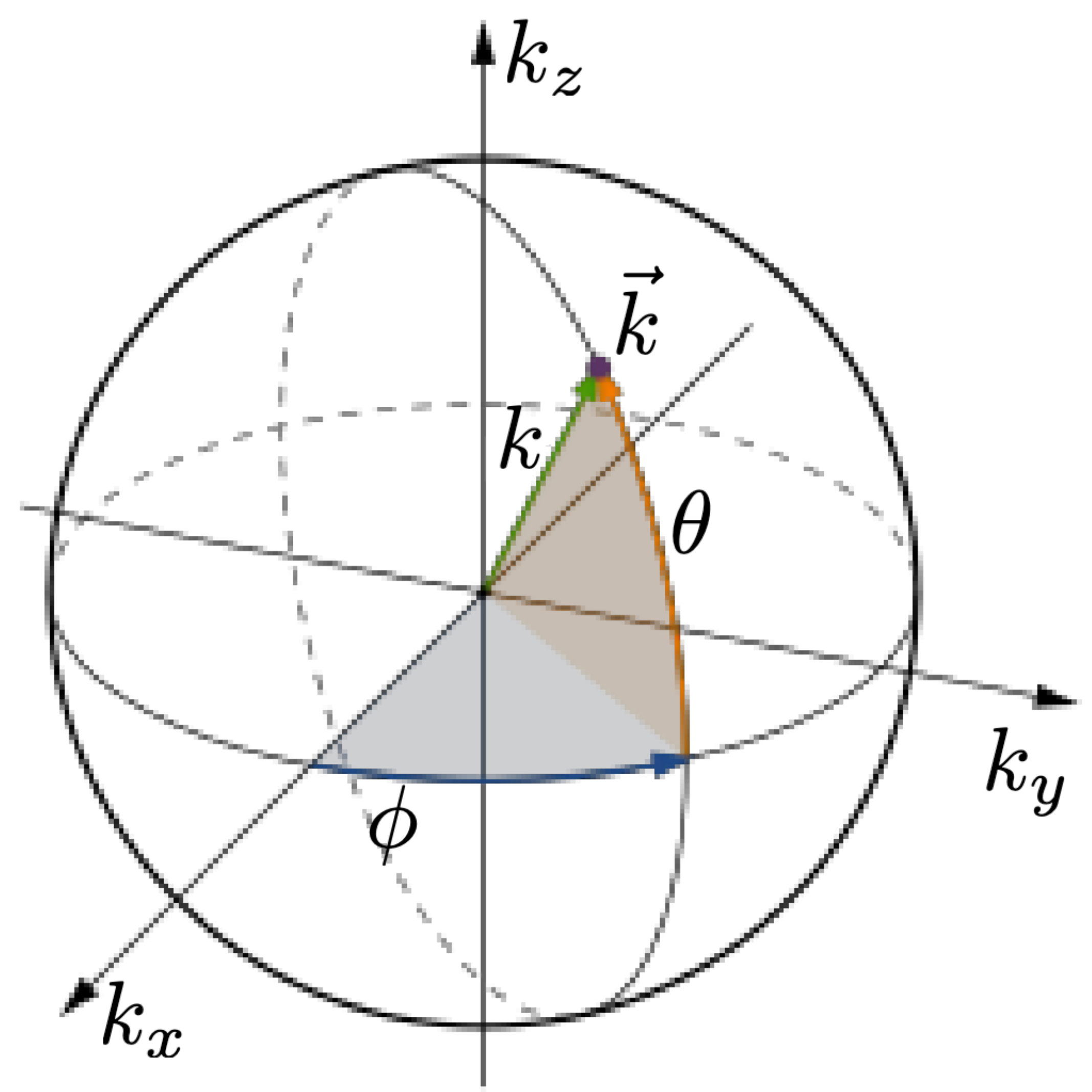}
     \includegraphics[width=0.49\textwidth,height=0.35\textheight, keepaspectratio]{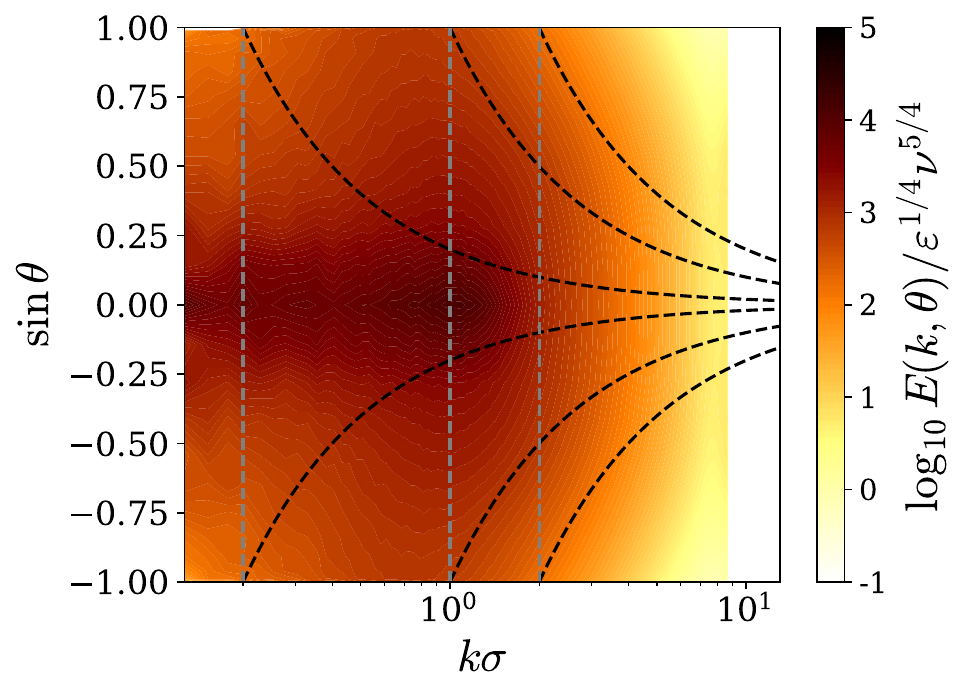}\\
     \includegraphics[width=0.49\textwidth,height=0.35\textheight, keepaspectratio]{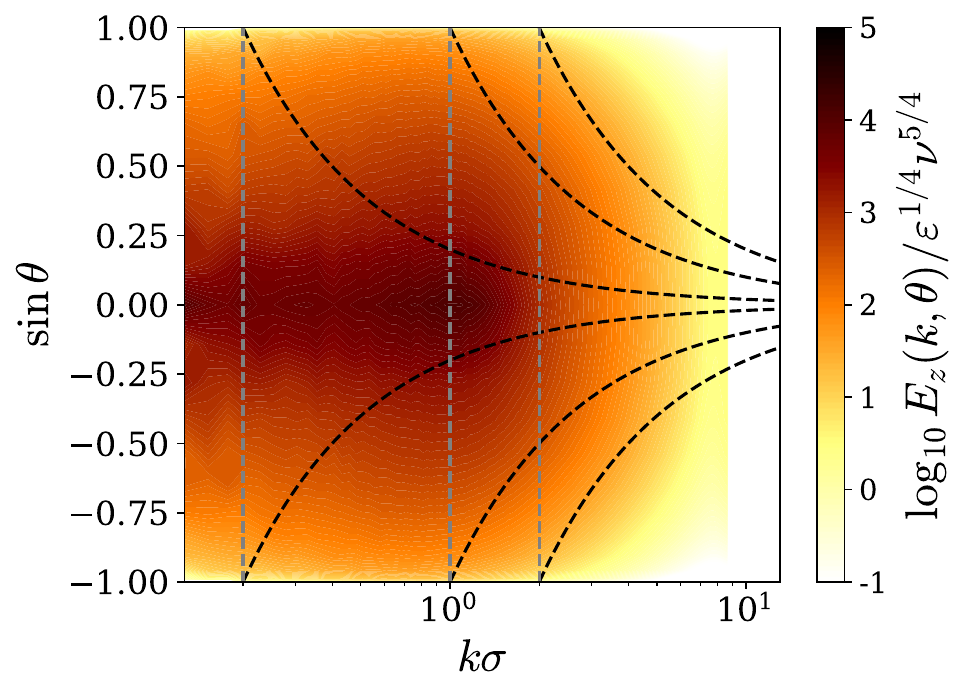}
     \includegraphics[width=0.49\textwidth,height=0.35\textheight, keepaspectratio]{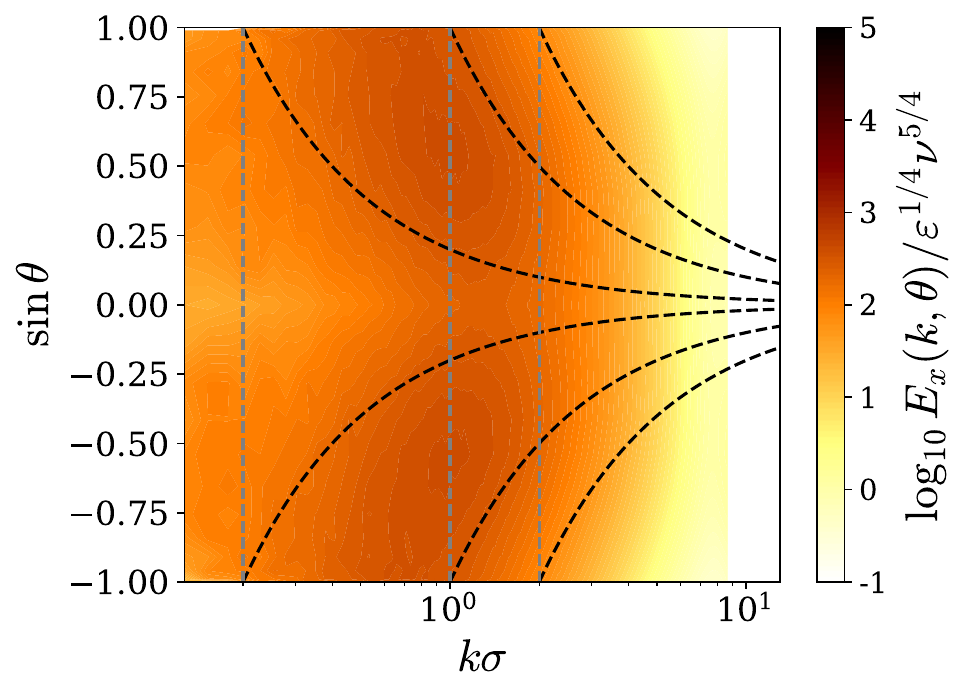}
     \caption
     {
	 (a) Scheme of spherical coordinates in the Fourier space.
	 (b) (c) and (d) Directional spectra of, respectively, the velocity vector, the vertical velocity component and a horizontal velocity component,  for $\alpha=10\%$.
	 Note that the spherical  spectra corresponds to an integral along a gray line, while the longitudinal spectra in the vertical direction are obtained by integration along a red line.
   	 }
   \label{fig:bubbles_spec_aniso}
   \end{figure}

To extend the discussion on the large scales, we need to take into account the anisotropy of the flow.
For the characterization of the anisotropy, it is necessary to distinguish the fact that the energy can be carried mainly by one component of the velocity vector (anisotropy between components) from the fact that the fluctuations in certain directions can carry more energy (directional anisotropy) \citep{Sagaut:2008}.
To characterize the latter, we consider a spherical coordinate system of the wave vector space, as schematized in figure \ref{fig:bubbles_spec_aniso}.
The angle $\theta$  characterizes the orientation of the wave vector with respect to the vertical direction: $\sin \theta = k_z / |k| $ ($\sin \theta =0$ corresponds to fluctuations in the horizontal direction and $\sin \theta =\pm 1$ to fluctuations in the vertical direction).
We then consider the directional spectra $E(k,\theta)$ which allows decomposing the energy of the fluctuations according to the wavelength and the orientation with respect to the vertical direction.
More specifically, $E(k, \theta)$ is defined by integration on all ``longitudes'' for a fixed ``latitude'' and modulus of the wave vector:   $E(k, \theta) = \int 1/2 \phi_{ii}(\bm{k}') \delta(|\bm{k}'|-k)\delta(k'_z/|\bm{k}'|-\sin\theta)d^3\bm{k}'  / 2\pi \cos \theta = \int_0^{2\pi}1/2 \phi_{ii}(k,\theta,\phi)k^2\sin \theta d \phi  / 2\pi \cos \theta$. 
The normalization factor $2\pi \cos \theta$ is introduced to correct the geometrical effect due to the fact that a band near the poles covers a less important surface than a band near the equator.

We show in figure \ref{fig:bubbles_spec_aniso} the directional spectra of the velocity for $\alpha=10\%$. 
It can be seen that at large scales ($k<1/d$) the energy is concentrated in the horizontal direction.
This concentration is characteristic of vertically aligned tubular structures \citep{Cambon:1989}, which can be seen on the flow visualization in figure \ref{fig:bubble_swarm_visu}.
At smaller scales ($k \gg 1/d$), one can see that the directional spectra become invariant with $\theta$ indicating that the directional anisotropy tends to vanish.

To characterize the anisotropy between components, we present as well in figure \ref{fig:bubbles_spec_aniso} the directional spectrum for both the vertical and the horizontal components of the velocity. Note that the axisymmetry of the flow impose $E(k,\theta) = E_z(k,\theta)+2 E_x(k,\theta)$.
At large scales, we notice that the spectra of the vertical velocity is very similar to that of the total kinetic energy, which indicates that at these scales the vertical component carries almost all the kinetic energy. 
This can be explained simply by the fact that the forcing due to the bubbles is essentially vertical.
We can also note that the horizontal component of the velocity presents a very weak directional anisotropy.
Finally, at small scales, we notice that the flow tends to become much more isotropic and presents both a decrease of the difference between the components and between the directions.
This indicates that at scales where the energy injection is zero, there is a redistribution between the components, which ensures a return to isotropy as $k$ increases. 

It should be noted that when considering the spectra with angular dependence for both the horizontal and vertical components, no $k^{-3}$ zone is distinguished. 
Thus, as noticed by \citet{Bellet:2006} for the decaying turbulence under strong rotation, the $k^{-3}$ region for the spherically averaged spectra results from an average between the different directions and the different components.

If the structure of the flow becomes locally isotropic at scales much smaller than $d$, one should expect the appearance of a $ k^{-5/3}$ inertial range for the velocity spectra, if the local Reynolds is large enough, since at these scales there is no more energy injection.
Such inertial range is not present in the numerical simulations reported here, but seems visible in the spectra obtained from the experiments of  \citet{Riboux:2010} (see figure \ref{fig:bubble_swarm_spectra1D_velocity} left).
In case the velocity spectra first present evolution as $E(k)\sim f^2k^{-3}$ followed by an inertial range  $E(k)\sim \langle \varepsilon\rangle^{2/3} k^{-5/3} $, the characteristic length of the crossover between these two regimes would be given by $\ell_I = \sqrt{\langle \varepsilon\rangle/f^3}$.
It is interesting to note that this length scale corresponds to  the classical estimates of the scale from which a turbulent flow subject to mean shear can be considered as locally isotropic \citep{Champagne:1970,Pope:2000}.
Taking the usual Kolmogorov scale $\eta=\nu^{3/4}\langle \varepsilon\rangle^{3/4}$, the extension of the inertial regime is given by $\ell_I/\eta=(\langle \varepsilon\rangle/\nu f^2)^{3/4}$.
In the simulations, as mentioned previously, no inertial range is present ($\ell_I=\eta$) and we have indeed $\langle \varepsilon\rangle = \nu f^2$.
It is likely that a larger power injection in the simulations (e.g. increasing $C_D$) would allow obtaining a separation between the scales of return to isotropy and the dissipative scales and thus to obtain a $-5/3$ range in agreement with the experiments.
Nevertheless, even in the absence of the $k^{-5/3}$ regime, this indicates that the rate of shear imposed by the bubble wakes control the relaxation to small-scale isotropy.

\section{Conclusion}\label{sec:conc}

Using a coarse-grained numerical approach, we have obtained Euler-Lagrange simulations of the flow agitation induced by a homogeneous swarm of rising bubbles at large Reynolds number. In this approach the momentum transferred from the bubbles to the fluid is filtered at a scale of the order of the bubble size and much larger than the mesh resolution.

Compared to DNS that are currently done with a comparable resolution, this method has the advantage that the approximation done in the description of the flow close to the bubble is explicit, and that the absence of interfacial jumps allows a straightforward computation of the spectrum of all quantities under interest. The main limitation is that the hydrodynamic force on the bubble relies on a model and is filtered at a scale close to the bubble diameter, which makes the technique poorly predictive regarding the bubble average velocity and the total energy injected by the bubble to the fluid. Comparisons with experiments indeed confirm this weakness, which impacts the predicted level of fluctuating energy.
However, the crucial mechanism of wake interactions is well reproduced, leading to the expected exponential decrease of the mean wake with the downstream distance. As a consequence, when normalized, the spectra of the velocity, both in frequency and wavenumber, are representative of real flows. In particular, with the present mesh-grid resolution and coarse-grained filtering, reliable spectra of all the terms of the energy budget are obtained in the range of wavenumbers extending from the peak of energy production by the bubbles up to the dissipative range. The examination of the spectral balance allows us to draw original conclusions which are summed up below.

The specificity of bubble-induced turbulence is the existence, in the spectral domain, of a subrange which begins when the production $P$ sharply decreases, and stops at the Kolmogorov dissipative scale. This regime, where the velocity spectrum broadly evolves as $k^{-3}$, is characterized by a constant shear-rate $f$ proportional to the average shear-rate of the bubble wakes. It constitutes a transition between the production dominated range, which scales as $f$ and the bubble diameter $d$, and the dissipative range, which scales as $f$ and the viscosity. It turns out that the $k^{-3}$ power-law can be obtained by asymptotically matching these two ranges. Unexpectedly, whereas $D$ evolves as $k^{-1}$, neither $P$ nor the spectral density transfer $T$ follows a power-law scaling. Consequently, even though the average production by buoyancy is balanced by the average dissipation $\epsilon$, the spectral dissipation density $D$ is not in equilibrium with the spectral production density $P$, which rules out the mechanism speculated by \cite{Lance:1991}. A return to isotropy is observed during this transitional regime, which led us to suggest a possible mechanism where the characteristic shear rate $f$ controls the rate of return to isotropy of the flow at small scales. 

These results shed a new light on the dynamic of bubble-induced turbulence, which is, however, not yet fully understood. In particular, we still ignore if increasing the Reynolds number could lead to a significantly wider $k^{-3}$ subrange. This question has no major practical implications since bubbly flows with significantly larger Reynolds numbers are not common. However, it is a fundamental issue for the comprehension of the underlying physical mechanism. The answer is left to future work with much more powerful computational means or smarter approaches.

\bigskip
\section*{Acknowledgement} 

This work was performed using HPC resources from GENCI-IDRIS and CALMIP center of the University of Toulouse. We thank the CEA Cadarache for financial support.

\bigskip 

For the purpose of Open Access, a CC-BY public copyright license has been applied by the authors to the present document and will be applied to all subsequent versions up to the Author Accepted Manuscript arising from this submission.

\bigskip 

Declaration of Interests. The authors report no conflict of interest.

\bibliographystyle{jfm}
\bibliography{biblio}

\end{document}